\documentclass[twocolumn]{aastex62}

\usepackage{ulem}
\usepackage{amsmath}

\graphicspath{{./}{figures/}}
\def \OIII {[O\,{\sc iii}]}

\usepackage{color}

\shorttitle{An Unusual {\it WISE} Flare}
\shortauthors{Yang et al.}

\begin{document}

\title{An Unusual Mid-Infrared Flare in a Type 2 AGN: An Obscured Turning-on AGN or Tidal Disruption Event?}

\correspondingauthor{Qian Yang}
\email{qiany@illinois.edu}

\author[0000-0002-6893-3742]{Qian Yang}
\affiliation{Department of Astronomy, University of Illinois at Urbana-Champaign, Urbana, IL 61801, USA}

\author[0000-0003-1659-7035]{Yue Shen}
\altaffiliation{Alfred P. Sloan Research Fellow}
\affiliation{Department of Astronomy, University of Illinois at Urbana-Champaign, Urbana, IL 61801, USA}
\affiliation{National Center for Supercomputing Applications, University of Illinois at Urbana-Champaign, Urbana, IL 61801, USA}

\author[0000-0003-0049-5210]{Xin Liu}
\affiliation{Department of Astronomy, University of Illinois at Urbana-Champaign, Urbana, IL 61801, USA}
\affiliation{National Center for Supercomputing Applications, University of Illinois at Urbana-Champaign, Urbana, IL 61801, USA}

\author{Xue-Bing Wu}
\affiliation{Department of Astronomy, School of Physics, Peking University, Beijing 100871, People's Republic of China}
\affiliation{Kavli Institute for Astronomy and Astrophysics, Peking University, Beijing 100871, People's Republic of China}

\author{Linhua Jiang}
\affiliation{Kavli Institute for Astronomy and Astrophysics, Peking University, Beijing 100871, People's Republic of China}

\author{Jinyi Shangguan}
\affiliation{Max-Planck-Institut f\"{u}r extraterrestrische Physik, Gie{\ss}enbachstr. 1, D-85748 Garching, Germany}

\author{Matthew J. Graham}
\affiliation{California Institute of Technology, 1200 E. California Blvd, Pasadena, CA 91125, USA}

\author{Su Yao}
\affiliation{Kavli Institute for Astronomy and Astrophysics, Peking University, Beijing 100871, People's Republic of China}


\begin{abstract}
We report the discovery of an exceptional MIR flare in a Type 2 AGN, SDSS J165726.81+234528.1, at $z=0.059$. This object brightened by 3 mag in the Wide-field Infrared Survey Explorer ({\it WISE}) $W1$ and $W2$ bands between 2015 and 2017 (and is fading since 2018), without significant changes ($\lesssim$ 0.2 mag) in the optical over the same period of time. Based on the {\it WISE} light curves and near-IR imaging, the flare is more significant at longer wavelengths, suggesting an origin of hot dust emission. The estimated black hole mass ($\sim 10^{6.5}\,M_\odot$) from different methods places its peak bolometric luminosity around the Eddington limit. The high luminosity of the MIR flare and its multi-year timescale suggest that it most likely originated from reprocessed dust radiation in an extended torus surrounding the AGN, instead of from stellar explosions. The MIR color variability is consistent with known changing-look AGN and tidal disruption events (TDEs), but inconsistent with normal supernovae. We suggest that it is a turning-on Type 2 AGN or TDE, where the optical variability is obscured by the dust torus during the transition. This MIR flare event reveals a population of dramatic nuclear transients that are missed in the optical.


\end{abstract}


\section{Introduction} \label{sec:introduction}

Modern multi-epoch and multi-wavelength data have enabled a broad range of time-domain studies from stellar transients to persistent variability from AGN. The multi-epoch \textit{Wide-field Infrared Survey Explorer} \citep[{\it WISE};][]{Wright2010} data in mid-infrared (MIR) $W1$ (3.4 $\mu$m) and $W2$ (4.6 $\mu$m) bands, in particular, can probe the changes in the continuum emission of warm/hot dust in different environments. MIR variability has been seen following optical transient/variability events, such as changing-look AGN \citep[CL AGN;][]{Sheng_etal_2017,Yang2018, Stern2018,Ross_etal_2018}, tidal disruption events \citep[TDE;][]{Blanchard2017, Jiang2017}, and supernova (SNe) explosion. In these cases, the MIR emission is predominantly thermal radiation from heated dust grains in the dust torus in AGN or from interstellar medium (ISM)/circumstellar medium (CSM) in SNe. In rare cases, non-thermal radiation from relativistic electrons accelerated by the shock waves in supernova remnants \citep{Dwek1987} or by jet launched in radio-loud AGN \citep{Urry1995} could also make contributions to the observed MIR emission.


The different classes of transients that produce both optical and MIR variability have characteristic spectral features. CL AGN are objects with emerging or disappearing broad emission lines accompanied by large-amplitude continuum variability, possibly caused by changes in the accretion of gas onto the central supermassive black hole \citep[SMBH;][]{LaMassa2015, Runnoe2016, Yang2018, MacLeod2016, MacLeod2019,Rumbaugh2018}. TDEs exhibit emission from helium and/or hydrogen, and some of them additionally show transient iron coronal lines when a star is  disrupted by the SMBH \citep{Komossa2008, Wang2012, vanVelzen2011, Gezari2012, Holoien2014, Arcavi2014, Komossa2015, Holoien2016, Holoien2018}. SNe show strong UV/optical brightening with or without hydrogen, silicon, and helium lines due to the core-collapse of a massive star or the thermonuclear explosion of a white dwarf accreting matter from a companion \citep[e.g.,][]{Filippenko1997, Hillebrandt2000, Heger2003}.

To date, most of these transient events are discovered by their variability from UV/optical or X-ray surveys and classified by follow-up spectroscopic characterization. However, transient events might be hidden within dusty environments and would not be detectable in optical, UV, or soft X-ray. For example, \citet{Mattila2018} reported a dust-enshrouded TDE discovered in the near-IR. The multi-epoch imaging from the {\it WISE} survey enables systematic discoveries of MIR transient/variable events, which can be cross-correlated with optical light curves. For example, \citet{Assef2018} identified 45 WISE-selected AGN candidates that are highly variable in the MIR using the AllWISE catalog; only seven of them show significant optical variability.



\begin{figure*}[htbp]
\hspace*{-0.5cm}
\epsscale{1}
\plotone{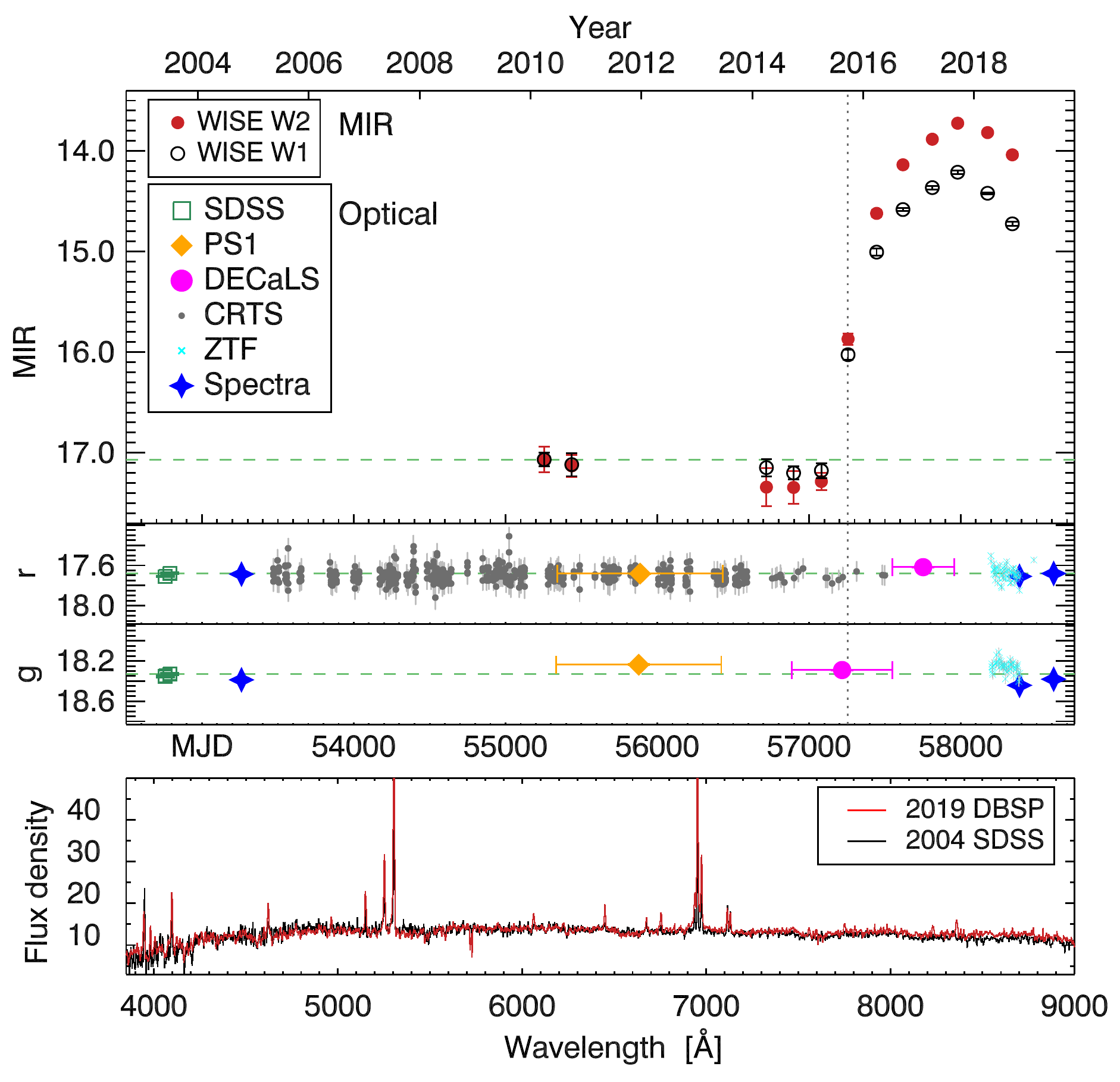}
\caption{\textbf{Top panel:} Light curves of J1657$+$2345 in MIR from {\it WISE} and in optical from various surveys, including SDSS, PS1, DECaLS, CRTS, and ZTF. All magnitudes are AB magnitudes. To compare the optical and MIR data, the y-axis in the three panels are plotted on the same scale. The horizontal dashed lines show the early epoch magnitude in $g$, $r$, and $W1$ bands from bottom to top panels. The vertical dotted line shows the first {\it WISE} epoch when the object began to brighten. J1657$+$2345 flared for more than 3 mag in MIR from 2015 to 2017. However, there is no significant variability in the optical from 2003 to 2019 (constant within 0.2 mag). \textbf{Bottom panel: } Spectra of J1657$+$2345 taken in 2019 by P200/DSBSP (red) and taken in 2004 by SDSS (black). There is no obvious variability between the two optical spectra.\label{fig:LC} }
\end{figure*}

Here we present an exceptional MIR flare in J$165726.81+234528.1$ (hereafter, J1657+2345) from our ongoing systematic study of {\it WISE} variability for AGN and galaxies. J1657+2345 is spectroscopically identified as a Type 2 AGN at $z = 0.059$ in SDSS \citep{York2000}. Its MIR light curves remained quiescent until the end of 2015, followed by an exceptionally large flare in both $W1$ and $W2$. Among $>1.8$ million spectroscopic galaxies in the SDSS fourteenth data release \citep[DR14,][]{Abolfathi2018}, J1657+2345 is identified as the most dramatic case with its {\it WISE} magnitudes brightened by 2.9 and 3.3 mag (a factor of 13.8 and 21.5 increase in flux) in $W1$ and $W2$ bands, respectively, between 2015 and 2017. According to the latest {\it WISE} data taken in August, 2018, J1657+2345 has past its peak MIR luminosity and is fading (see Fig.\ \ref{fig:LC}). 

We compiled all available optical photometric data from various surveys, including the SDSS, Pan-STARRS \citep[PS1,][]{Chambers2016}, the Dark Energy Camera Legacy Survey \citep[DECaLS,][]{Dey2019}, the Catalina Real-time Transient Survey \citep[CRTS,][]{Drake2009}, and the Zwicky Transient Facility \citep[ZTF,][]{Bellm2019}. The CRTS data 8 months after the onset of the MIR flare shows that it was still quiescent in the optical. To identify potential spectral variability, we obtained new optical spectra in September, 2018 and May, 2019. There is no significant difference between the new spectra and the earlier SDSS spectrum taken in 2004 before the MIR flare. The optical photometric data constrain a maximum variability of $\sim 0.2$ mag before and after the onset of the MIR flare. Thus this object is markedly different from any previous transients first identified in the optical. It is reminiscent of the handful of MIR variable AGN candidates with low optical variability reported in \citet{Assef2018}, but the contrast of the MIR and optical variability is much more extreme. The large difference between the MIR and optical variability motivates a thorough investigation of this event to explore possible scenarios on the nature of the extreme MIR-only flare. 

In \S\ref{sec:data}, we describe the observations of J1657$+$2345 in MIR, optical, and near-infrared. We describe the variability, spectral energy distribution (SED), and spectral features of J1657$+$2345 in \S\ref{sec:results}. In \S\ref{sec:disscussion}, we compare its MIR variability to CL AGN, TDEs, and SNe. We discuss the timescales of MIR variability from reprocessing the optical/UV variability with simple geometric dust torus models. We conclude in \S\ref{sec:summary}. In this paper, we use a $\Lambda$CDM cosmology with parameters $\Omega_{\Lambda}=0.7$, $\Omega_{\rm m}=0.3$, and $H_0=70$ km s$^{-1}$ Mpc$^{-1}$. Unless otherwise specified, all magnitudes are in the AB system \citep{Oke1983}. 

\section{Observations and Data} \label{sec:data}

\begin{deluxetable*}{lcccccccccc}
\tablecaption{{\it WISE} MIR Photometry \label{tab:MIR}}
\tablewidth{1pt}
\tablehead{
\colhead{Date} &
\colhead{MJD} &
\colhead{Number} &
\colhead{$W1$} &
\colhead{$W2$} &
\colhead{$W3$} &
\colhead{$W4$}
} 
\startdata
2010 Feb 28 & 55255 & 15 & 17.07 (0.07) & 17.07 (0.13) & 15.36 (0.16) & 14.00 (0.28)\\
2010 Aug 28 & 55436 & 17 & 17.12 (0.11) & 17.13 (0.11) & $\cdots$ & $\cdots$ \\
2014 Mar 3 & 56719 & 14 & 17.15 (0.09) & 17.34 (0.19) & $\cdots$ & $\cdots$ \\
2014 Aug 29 & 56898 & 15 & 17.20 (0.07) & 17.34 (0.17) & $\cdots$ & $\cdots$ \\
2015 Feb 28 & 57081 & 14 & 17.18 (0.07) & 17.29 (0.09) & $\cdots$ & $\cdots$ \\
2015 Aug 23 & 57257 & 17 & 16.03 (0.05) & 15.87 (0.06) & $\cdots$ & $\cdots$ \\
2016 Feb 27 & 57445 & 15 & 15.01 (0.04) & 14.62 (0.03) & $\cdots$ & $\cdots$ \\
2016 Aug 19 & 57619 & 14 & 14.58 (0.01) & 14.14 (0.02) & $\cdots$ & $\cdots$ \\
2017 Feb 28 & 57812 & 17 & 14.37 (0.02) & 13.88 (0.02) & $\cdots$ & $\cdots$ \\
2017 Aug 14 & 57979 & 9 & 14.21 (0.02) & 13.73 (0.03) & $\cdots$ & $\cdots$ \\
2018 Feb 28 & 58177 & 15 & 14.42 (0.01) & 13.82 (0.02) & $\cdots$ & $\cdots$ \\
2018 Aug 10 & 58340 & 9 & 14.72 (0.02) & 14.04 (0.03) & $\cdots$ & $\cdots$ \\
\enddata
\tablecomments{The {\it WISE} magnitudes are converted to AB magnitudes. The values in the parentheses are the magnitude errors.}
\end{deluxetable*}

\subsection{{\it WISE} Photometry}

{\it WISE} scanned the full sky from January to July in 2010 in four bands centered at wavelengths of 3.4, 4.6, 12, and 22 $\mu$m ($W1$, $W2$, $W3$, and $W4$). The secondary cryogen survey and Near-Earth Object Wide-field Infrared Survey Explorer \citep[{\it NEOWISE};][]{Mainzer2011} Post-Cryogenic Mission mapped the sky from August, 2010 to February, 2011. The {\it NEOWISE} Reactivation Mission \citep[{\it NEOWISE-R};][]{Mainzer2014} surveys the sky in $W1$ and $W2$ bands from 2013 twice a year. {\it WISE} obtains $\sim 10-20$ observations within a 36-hrs window in each visit. We calculate the median magnitude and magnitude error, specifically the semi-amplitude of the range enclosing the 16th and 84th percentiles of all flux measurements within a 6-month window (summarized in Table \ref{tab:MIR}). We limit to good quality single-epoch data points with the best frame image quality score ($qi\_fact=1$), observed far away from the South Atlantic Anomaly ($saa\_sep \geq 5$), with no contamination from the moon ($moon\_masked=0$), and excluding spurious detection ($cc\_flags=0$). The {\it WISE} magnitudes are converted from Vega to AB magnitude as $m_{\rm AB} = m_{\rm Vega} + \Delta m$, where $\Delta m$ is 2.699, 3.339, 5.174, and 6.620 in $W1$, $W2$, $W3$, and $W4$ bands, respectively. 

\begin{deluxetable*}{lcccccc}
\tablecaption{\ Optical Photometry \label{tab:Optical}}
\tablewidth{1pt}
\tablehead{
\colhead{Survey} &
\colhead{Photometry} &
\colhead{MJD} &
\colhead{Year} &
\colhead{$g$} &
\colhead{$r$} &
\colhead{$z$}
}
\startdata
SDSS & Model & 52788 & 2003 & 18.33 (0.01) & 17.68 (0.01) & 17.05 (0.02)\\
\hline
PS1 & Kron & 55333-56422 & 2010-2013 & 18.24 (0.01) & $\cdots$ & $\cdots$ \\
 & & 55341-56433 & 2010-2013 & $\cdots$ & 17.68 (0.01) & $\cdots$ \\
 & & 55275-56520 & 2010-2013 & $\cdots$ & $\cdots$ & 17.06 (0.01) \\
\hline
DECaLS & Model & 56888-57548 & 2014-2016 & 18.29 (0.01) & $\cdots$ & $\cdots$ \\
 & & 57548-57956 & 2016-2017 & $\cdots$ & 17.62 (0.01) & $\cdots$\\
 & & 57110-57120 & 2015 & $\cdots$ & $\cdots$ & 17.02 (0.01) \\
\hline
ZTF & Aperture & 58204-58389 & 2018 & 18.26 (0.04) & $\cdots$ & $\cdots$ \\
 & & 58198-58482 & 2018 & $\cdots$ & 17.68 (0.03) & $\cdots$ \\
\hline
CRTS & Aperture(unfiltered) & 53474-57500 & 2005-2016 & $\cdots$ & 17.68 (0.10) & $\cdots$ \\
\hline
SDSS/spec & Spectrophotometry & 53260 & 2004 & 18.39 (0.01) & 17.69 (0.01) & 17.12 (0.01) \\
\hline
2.16m  & Spectrophotometry & 58387 & 2018 & 18.44 (0.01) & 17.71 (0.01) & 17.26 (0.06) \\
\hline
DBSP & Spectrophotometry & 58612 & 2019 & 18.38 (0.01) & 17.68 (0.01) & 17.02 (0.01)\\
\enddata
\tablecomments{The photometry of PS1 and DECaLS is from stacked images. The magnitudes of ZTF and CRTS in this table are median magnitudes. The compiled light curves show that there is no optical variability (more than 0.2 mag) from 2003 to 2019.}
\end{deluxetable*}

\subsection{Optical Photometry}
We compile all available optical photometric data from various surveys, including SDSS, PS1, DECaLS, CRTS, and ZTF (see Table \ref{tab:Optical}). J1657$+$2345 was observed in the SDSS imaging survey in $ugriz_{\rm SDSS}$ bands in May, 2003. As J1657$+$2345 is an extended source at $z=0.059$, we use the SDSS model mag, obtained by fitting to de Vaucouleurs (elliptical galaxies) or exponential (spiral galaxies) models to the photometric data. The SDSS $gri$ magnitudes are nearly AB. The SDSS $u$-band and $z$-band magnitudes are corrected to the AB system using $u_{\rm AB} = u_{\rm SDSS} - 0.04$ mag and $z_{\rm AB} = z_{\rm SDSS} + 0.02$ mag \citep{Fukugita1996}. J1657$+$2345 was observed by PS1 \citep{Chambers2016} from March, 2010 to June, 2014 in $grizy_{\rm PS1}$ bands. We use the PS1 \citet{Kron1980} magnitude from the PS1 stack catalog. J1657$+$2345 was observed three times in each band by DECaLS from August, 2014 to June, 2016 in $g_{\rm DECAM}$ band (3 epochs), from June, 2016 to July, 2017 in $r_{\rm DECAM}$ band (3 epochs), and from March to April, 2015 in $z_{\rm DECAM}$ band (3 epochs). These images were stacked and presented in the DECaLS DR7 catalog \citep{Dey2019}. We use the DECaLS model magnitude, obtained by fitting five morphological types including point sources, round exponential galaxies with a variable radius, de Vaucouleurs profiles, exponential profiles, or composite profiles. The Catalina Real-time Transient Survey \citep[CRTS,][]{Drake2009} repeatedly observed a large portion of the sky. J1657$+$2345
was observed more than 450 times by CRTS from 2005 to 2016. The CRTS data are aperture-based, unfiltered photometry. The Zwicky Transient Facility \citep[ZTF;][]{Bellm2019} is a new time-domain survey from 2017 in $gri_{\rm ZTF}$ bands. J1657$+$2345 was covered by ZTF about 50 times in $g$ and $r$ bands from March to December in 2018. The ZTF data are aperture-based photometry with a typical aperture diameter of 2\arcsec. 

\subsection{Optical Photometric Data Calibration}

To calibrate the optical data from different surveys onto the same flux scale, we apply additive corrections to the optical magnitudes taking into account different filter curves and photometry methods. We convert all magnitudes to the AB system. To correct for different filter curves, we convolve the DBSP spectrum (see Section \ref{sec:spec}) with the PS1/DECaLS/ZTF filter curves to obtain synthetic magnitudes, and compare to those derived with the SDSS filters to derive the corrections. Thus the calibration offsets for PS1 are 0.09, 0.02, and $-0.06$ mag in $grz$ bands; for DECaLS are 0.07, 0.13, and 0.04 mag in $grz$ bands; and for ZTF are 0.06 and 0.13 mag in $gr$ bands. We apply an additional correction for ZTF magnitudes as ZTF uses aperture-based photometry. According to the DECaLS photometry, the offsets between model magnitude and 2\arcsec aperture magnitude for J1657$+$2345 are $-0.50$ and $-0.46$ mag in $g$ and $r$ bands, respectively. Therefore, ZTF magnitudes are further corrected by $-0.44$ and $-0.33$ mag in $g$ and $r$ bands. CRTS data are aperture-based photometry and observed unfiltered, so we apply a constant offset $+0.18$ to the CRTS magnitudes to match the median CRTS magnitude (18.50 mag) to the contemporary calibrated PS1 $r$-band magnitude (18.68 mag). We summarize the calibrated $grz$ photometry in Table \ref{tab:Optical}.

\subsection{Spectroscopic Observations \label{sec:spec}} 

J1657$+$2345 was observed by SDSS on September 12, 2004. The SDSS spectroscopy covers a wavelength range from 3820 to 9185 ${\rm \AA}$ with a spectral resolution of $R = \lambda/\Delta \lambda \sim 2000$ \citep{Abazajian2009} and a spectral binning of 69 km s$^{-1}$ per pixel. The median signal-to-noise ratio (S/N) per pixel of the SDSS spectrum for J1657$+$2345 is 15. J1657$+$2345 has no obvious broad emission lines in the spectrum. The narrow emission line flux ratios suggest it is photoionized by AGN, with ${\rm log([O~III/H\beta)}=0.65\pm0.01$ and ${\rm log([N~II]/H\alpha)}=-0.24\pm0.01$ \citep{Kauffmann2003}. Therefore, J1657$+$2345 is a Type 2 AGN, specifically, a Seyfert 2 galaxy according to the division line between Seyferts and LINERs \citep{CidFernandes2010}. The stellar velocity dispersion $\sigma_*$ is $60.8\pm11.2$ km s$^{-1}$, measured from the SDSS spectrum \citep{Thomas2013}. Using the local relation between SMBH mass $M_{\rm BH}$ and $\sigma_*$ \citep{Kormendy2013}, we estimate a BH mass of $10^{6.2\pm0.2}~M_\odot$. The measured $\sigma_*$ may be unreliable given SDSS's spectral resolution, therefore we will use different approaches to cross-check the BH mass estimate (\S\ref{sec:results}). 

We obtained an optical spectrum using the Xinglong 2.16 m telescope in China on September 26, 2018. We use the Beijing Faint Object Spectrograph and Camera (BFOSC) with Grism 4. The object was observed under seeing $\sim$ 2\arcsec, so we used a slit width of 2.3\arcsec. This instrument configuration yields a dispersion of 198 ${\rm \AA}$/mm, a wavelength coverage from 3850 to 8860 ${\rm \AA}$, and a resolution of $R\sim265$ \citep{Fan2016}. The object was observed with one exposure of 1800 seconds. We observed a standard star with the same configuration, HD 161817, for flux calibration. The spectrum was reduced using standard IRAF routines \citep{Tody1986, Tody1993}. The median S/N per pixel of the 2.16m spectrum is 6. 

We obtained another optical spectrum using the Palomar P200/DBSP spectrograph on May 9, 2019. We used DBSP with the G600 grating on the blue side with a central wavelength of 4000 ${\rm \AA}$ and G316 grating on the red side with a central wavelength of 7500 ${\rm \AA}$. The object was observed with a 1.5\arcsec slit under seeing $\sim 1.2\arcsec$. This configuration yields a dispersion of 71 ${\rm \AA}$/mm and a resolution of $R\sim969$ at 4000 ${\rm \AA}$ on the blue side; a dispersion of 135 ${\rm \AA}$/mm and a resolution of $R\sim958$ at 7500 ${\rm \AA}$ on the red side. We obtained one exposure of 900 seconds. We observed a standard star, BD$+$28d4211, for flux calibration. The DBSP spectrum covers a wide wavelength range, and the median S/N per pixel from 3800 to 9200 ${\rm \AA}$ is 9.

The spectra were taken in smaller aperture/slit than photometry, and J1657$+$2345 is an extended source. To correct for aperture loss, we calculate spectrophotometry by convolving the spectra with SDSS $grz$ filter curves, and compare the spectrophotometry with photometry. We applied a constant scaling factor to the SDSS spectrum, specifically a factor of 1.95, to match the SDSS spectrophotometry in $r$ band to SDSS $r-$band model photometry. We apply the same scaling factor to other spectra as their \OIII\ fluxes are consistent with that of the SDSS spectrum (see Section \ref{sec:spec_prop}).




\begin{deluxetable*}{lcccccccccc}
\tablecaption{Near-infrared Photometry \label{tab:NIR}}
\tablewidth{1pt}
\tablehead{
\colhead{Instrument} &
\colhead{Date} &
\colhead{MJD} &
\colhead{$Y$} &
\colhead{$J$} &
\colhead{$H$} &
\colhead{$K$}
} 
\startdata
2MASS & 2002 Feb 12 & 52317 & $\cdots$ & 16.85 (0.10) & 16.68 (0.17) & 16.40 (0.13)\\
UKIRT & 2019 Feb 26 & 58540 & 16.99 (0.02) & 16.72 (0.01) & 16.40 (0.01) & 16.02 (0.01)\\
\enddata
\tablecomments{All magnitudes are converted to AB. }
\end{deluxetable*}

\subsection{Near-Infrared Photometry}
J1657$+$2345 was observed by the Two Micron All Sky Survey 
\citep[2MASS;][]{Skrutskie2006} on February 12, 2002 in $J$, $H$, and $Ks$ bands. 
We downloaded the $J$, $H$, and $K_s$ band images from NASA/IPAC Infrared Science 
Archive (IRSA)\footnote{\url{irsa.ipac.caltech.edu/}}, and performed aperture 
photometry on the images using the Python package 
\texttt{photutils}.\footnote{http://photutils.readthedocs.io/}  The source-free 
background is fitted and subtracted with a two-dimensional 3rd-order polynomial 
function, which is flexible to fit the large-scale background gradient but 
robust not to fit the small-scale variation due to the source.  The 
flux of the source is integrated over a circular aperture of 4\arcsec\ in radius, and the 
local background is measured and removed with an annulus of radii between 25\arcsec\ and 
35\arcsec. By increasing the aperture radius till 12\arcsec, we find that the 
4\arcsec\ aperture size is large enough to enclose more than 95\% of the total flux. No aperture correction is applied.  
Our new measurements are systematically brighter than the 2MASS point source catalog results, by 0.29, 0.32, and 0.45 magnitude, respectively for $J$, $H$, 
and $K_s$ bands. However, our results are more consistent with the optical spectra and match the stellar emission model better (see Section 
\ref{sec:sed}). We converted the 2MASS Vega magnitudes to AB magnitude as 
$m_{\rm AB} = m_{\rm Vega} + \Delta m$, where $\Delta m$ is 0.89, 1.37, and 1.84 
in $J$, $H$, and $K_s$ bands.

We obtained new near-infrared imaging using the United Kingdom Infrared Telescope (UKIRT) on February 26, 2019 in $YJHK$ bands. We used the UKIRT Wide Field Camera (WFCAM) with a 4-point dithering pattern in $Y$ band, and 8-point dithering pattern in $JHK$ bands. At each pointing the exposure time was 10 seconds. WFCAM data were processed by the Cambridge Astronomical Survey Unit. We use UKIRT photometry derived from a 4\arcsec radius aperture for comparison with the 2MASS photometry. We converted the UKIRT Vega magnitudes to AB magnitude as $m_{\rm AB} = m_{\rm Vega} + \Delta m$, where $\Delta m$ is 0.634, 0.938, 1.379, and 1.900 in $YJHK$ bands \citep{Hewett2006}.

\begin{figure*}[htbp]
\hspace*{-0.5cm}
\epsscale{1.0}
\plotone{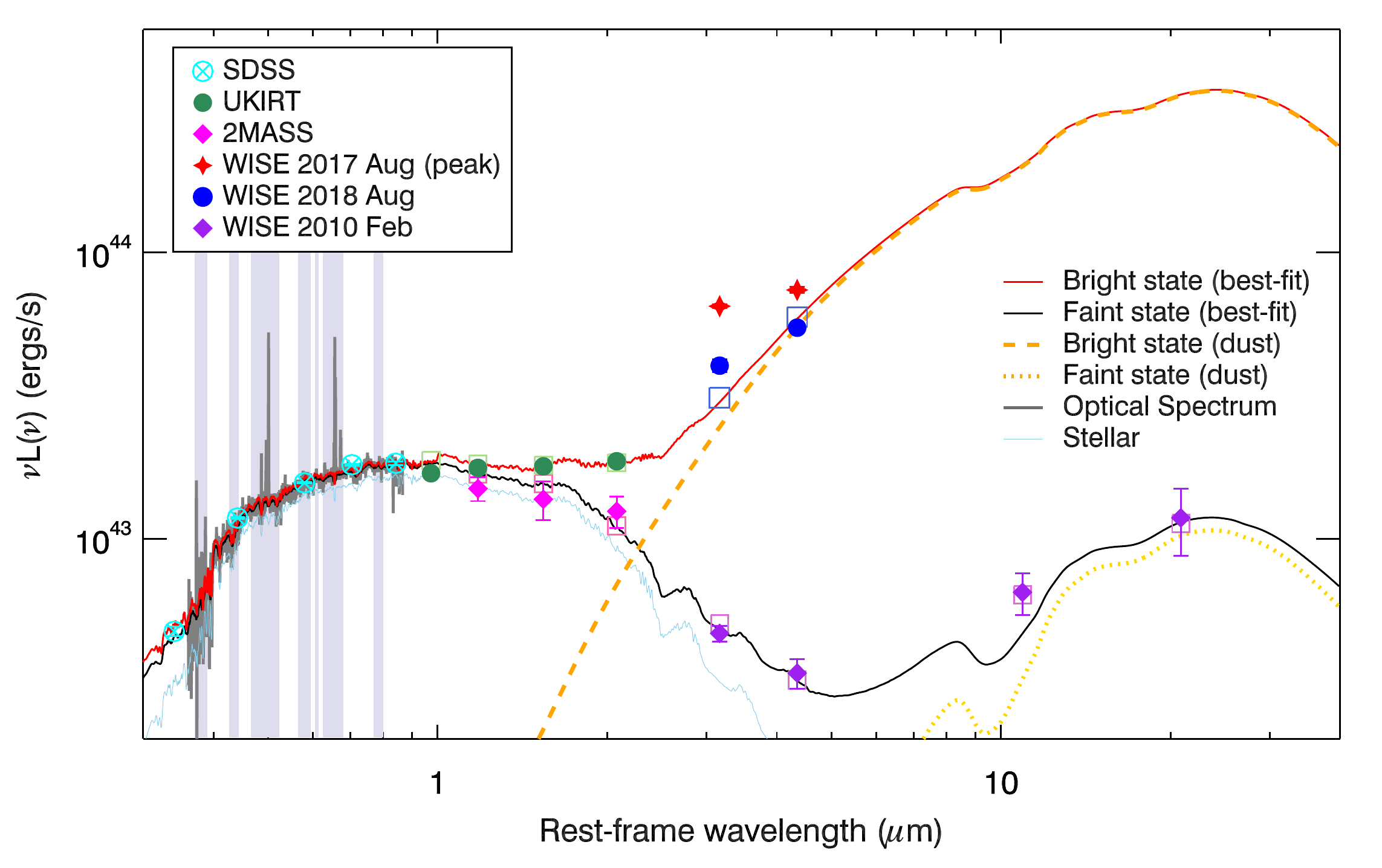}
\caption{SEDs of J1657$+$2345 in the faint state (black) and the bright state (red). We simultaneously fit the spectral and photometric data. The wavelength ranges with strong emission lines are excluded in the fit (shaded regions). We consider stellar emission (light-blue solid line), a power-law continuum (negligible), and a clumpy dust radiative transfer model (orange dashed line in bright state, yellow dotted line in faint state). The open squares are the modeled photometric data at the observed bands. The red stars are {\it WISE} data in August, 2017, which is the brightest {\it WISE} epoch. J1657$+$2345 varies more at redder wavelength. The hot dust enhanced a lot, while the optical light is not variable.
}
\label{fig:SED}
\end{figure*}

\begin{figure*}[htbp]
\hspace*{-0.5cm}
\epsscale{1}
\plotone{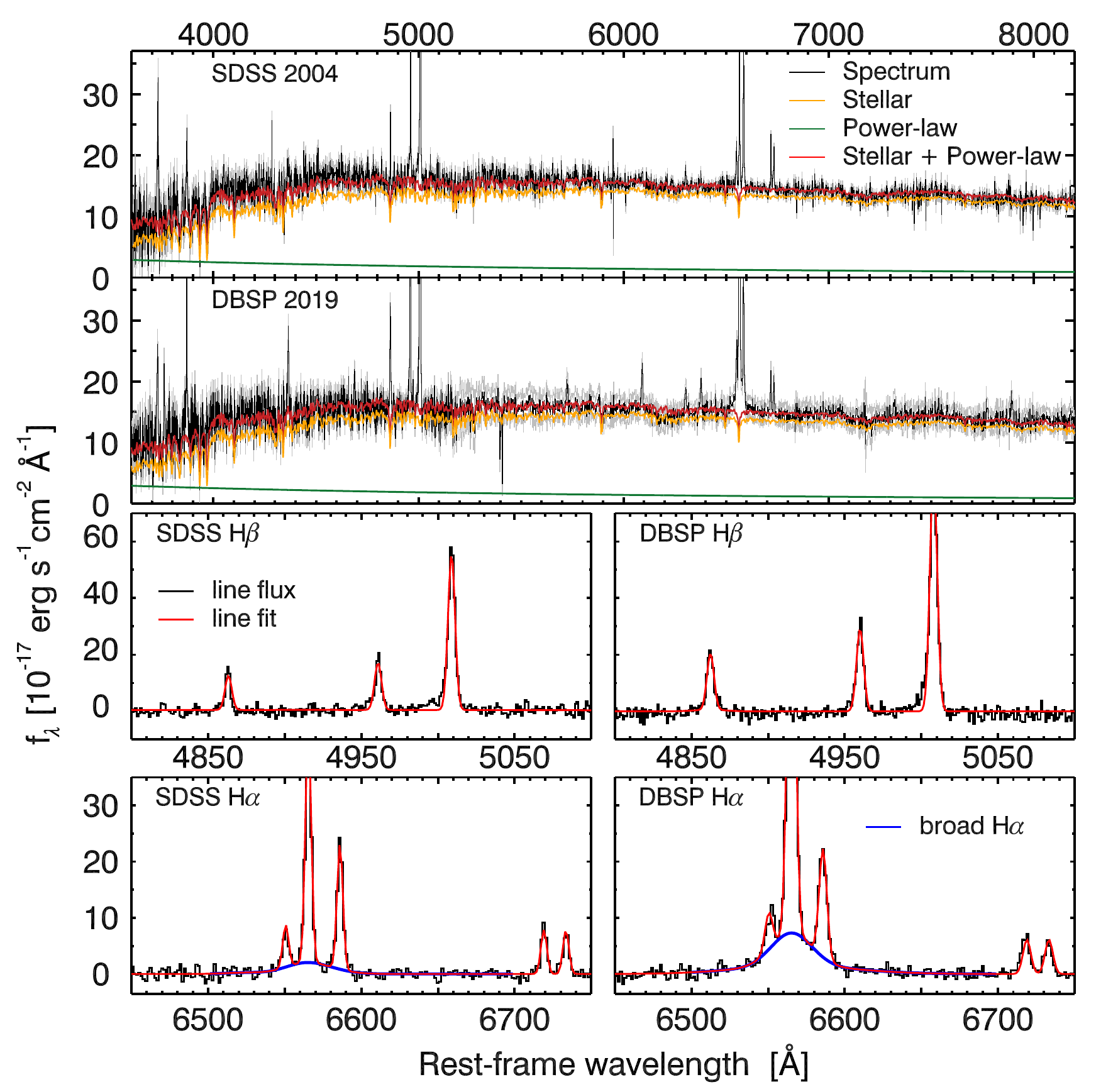}
\caption{Spectral fits to the SDSS and DBSP spectrum. \textbf{Top two panels:} The data (black), uncertainties (grey), stellar (orange), power-law continuum (green), and total continuum (red). The two spectra are well fitted by the same continuum components. The power-law (AGN) contribution is weak compared to host stellar emission. \textbf{Bottom four panels:} zoom-in for the H$\beta$ and H$\alpha$ fitting to the SDSS (left) and DBSP (right) spectra. Weak broad H$\alpha$ is detected in both spectra, which is consistent with scattered broad-line flux in Type 2 AGN (see text). The DBSP spectrum shows stronger broad H$\alpha$ emission line (blue). No broad H$\beta$ emission is detected in both spectra. \label{fig:spec} }
\end{figure*}

\section{Results} \label{sec:results}
\subsection{Variability of J1657$+$2345} 

Fig.\ \ref{fig:LC} displays the multi-wavelength light curves for J1657$+$2345 from our collected data. From 2010 to 2015, J1657$+$2345 was scanned by {\it WISE} for five times, during which it remained in the faint state. August 23, 2015 was the first epoch when {\it WISE} captured the flare. It flared by 1.15 and 1.42 mag in $W1$ and $W2$ bands within half a year from February to August 2015. It continued to brighten from 2015 to 2017. The peak-luminosity epoch caught by {\it WISE} is August 14, 2017, with $W1=14.21$ and $W2=13.73$ (AB magnitude). It brightened by 2.97 and 3.56 (a factor of 13.8 and 21.5 flux increase) in $W1$ and $W2$ bands during 2.5 years. The {\it WISE} observation on February 28, 2018 indicates that it started to fade. It was still in the bright state at the latest public {\it WISE} epoch (August 10, 2018), 2.46 and 3.25 mags brighter than its faint state, in $W1$ and $W2$ bands. 


However, there is no significant variability in the optical from 2003 to 2019. The optical photometric data from various surveys (summarized in Table \ref{tab:Optical}) are consistent with each other within $\sim 0.2$ mag. The continuous CRTS data from 2005 to 2016 is constant with a standard deviation of 0.07 mag. The stacked DECaLS $r$-band photometry from June 9, 2016 to July 22, 2017 is consistent with the SDSS and PS1 photometry within 0.1 mag. The ZTF data from March to December 2018 is also consistent with the SDSS, PS1, and DECaLS photometry within 0.1 mag. Furthermore, We find no evidence for significant flux variations between the SDSS spectrum, the 2.16m telescope spectrum, and the DBSP spectrum that were taken $\sim 15$ years apart. There is no continuum flux enhancement in the optical spectra compared to its earlier SDSS spectrum (we discuss the details on spectra in Section \ref{sec:spec_prop}). 

\begin{deluxetable*}{lll}
\tablecaption{Physical Properties of J1657+2345 \label{tab:fit}}
\tablewidth{1pt}
\tablehead{
\colhead{Property} &
\colhead{Value} &
\colhead{Note}
} 
\startdata
${\rm log}(M_{\rm *, old}~[M_\odot])$ & $9.59\pm0.02$ & Faint-state SED\\
${\rm log}(M_{\rm *, young}~[M_\odot])$ & $7.72\pm0.22$ & Faint-state SED\\
Age$_{\rm *, old}$ [Gyr] & $1.45\pm0.04$ & Faint-state SED\\
Age$_{\rm *, young}$ [Gyr] & $0.27\pm0.05$ & Faint-state SED\\
\hline
$L_{\rm bol,faint}~[\mathrm{erg\,s^{-1}}]$ & $(1.4\pm0.9)
\times10^{43}$ & Faint-state SED\\
& $(7.57\pm0.01)
\times10^{42}$ & SDSS $L_{[O III]}$ \\
& $(1.35\pm0.01)
\times10^{43}$ & DBSP $L_{[O III]}$ \\
\hline
$L_{\rm bol,bright}~[\mathrm{erg\,s^{-1}}]$ & $(4.9\pm0.1)
\times10^{44}$ & Bright-state SED\\
\hline
$L_{\rm bol,peak}~[\mathrm{erg\,s^{-1}}]$ & $(7.5\pm0.1)
\times10^{44}$ & {\it WISE} 2017 August\\
\hline
${\rm log}(L_{\rm [OIII]}~[\mathrm{erg\,s^{-1}}])$ & ${40.45\pm0.01}$ & SDSS spectrum\\
 &  ${40.66\pm0.01}$ & DBSP spectrum\\
\hline
${\rm log}(L_{\rm H\alpha,broad}~[\mathrm{erg\,s^{-1}}])$ & ${39.86\pm0.06}$ & SDSS spectrum \\
 & ${40.53\pm0.05}$ & DBSP spectrum\\
\hline
${\rm FWHM}_{\rm H\alpha, broad} [\mathrm{km\,h^{-1}}]$ & $1610\pm210$ & SDSS spectrum \\
 & $1664\pm161$ & DBSP spectrum\\
\hline
${\rm log}(M_{\rm BH}~[M_\odot])$  & ${6.22\pm0.28}$ & SDSS $\sigma_*$ \citep{Kormendy2013}\\
& $5.97\pm0.30$ & Stellar mass \citep{Reines2015}\\
& ${6.66\pm0.15}$ & SDSS H$\alpha$ scattered light \citep{Greene2005}\\
& ${7.05\pm0.10}$ & DBSP H$\alpha$ scattered light \citep{Greene2005}\\
\enddata
\tablecomments{All uncertainties are statistical errors only. }
\end{deluxetable*}

\subsection{SED Fitting} \label{sec:sed}

J1657$+$2345 brightened by 0.6 mag more in the redder $W2$ band than in $W1$ band, a factor of 1.6 more in flux. It brightened by 0.13, 0.28, and 0.38 mag in $J$, $H$, and $K$ bands, respectively, comparing the 2MASS and UKIRT photometry. Therefore we confirmed that the J1657$+$2345 flare is more prominent at longer IR wavelengths. 

We construct SEDs well before (faint state) and after (bright state) the onset of the flare. We simultaneously fit the optical spectrum and IR photometric data for each state. In the faint state, we use the SDSS spectrum (September, 2004), 2MASS photometry (February, 2002), and {\it WISE} data (February, 2010). In the bright state, we use the DBSP spectrum (May, 2019), UKIRT photometry (February, 2019), and the latest {\it WISE} data (August, 2018). 

Figure \ref{fig:SED} shows the SED fitting results in both states. To match the IR photometry with larger apertures, the spectrum is scaled by a factor of 1.95 to match the SDSS model mag (cyan). We perform the SED fitting with a Markov chain Monte Carlo (MCMC) method \citep{Shangguan2018}. We adopt the host galaxy stellar emission model that consists of two simple stellar population models \citep{Bruzual2003} with the \cite{Chabrier2003} initial mass function. The young stellar population has an age less than 300 Myr and the old stellar population has an age of 0.3--15 Gyr. The stellar masses and ages are free parameters in the fitting.  We incorporate a power-law component to fit the scattered UV emission from the accretion disk, which is found necessary for Type 2 AGN \citep{Bessiere2017,Zhao2019}.  The amplitude and the slope of the power-law model are free parameters in the fitting. A clumpy dust torus model \citep{Honig2017} is used to mainly fit the MIR data. The inclination angle, power-law index of the cloud radial distribution, number of clouds on the equatorial plane, the vertical scale height, and the total luminosity of the torus are fitted. Since we do not have complete MIR coverage to fit the torus component, we did not use more complicated models that include a wind component \citep{Shangguan2019}.  For the spectrum we only use segments of continuum without strong line emission.\footnote{We used the following continuum wavelength windows: 2900--3100, 3500--3700, 3900--4260, 4430--4660, 5240--5650, 5950--6050, 6150--6250, 6800--7700, 8000--9000 \AA.}

\begin{figure*}[htbp]
\hspace*{-0.5cm}
\epsscale{1.1}
\plotone{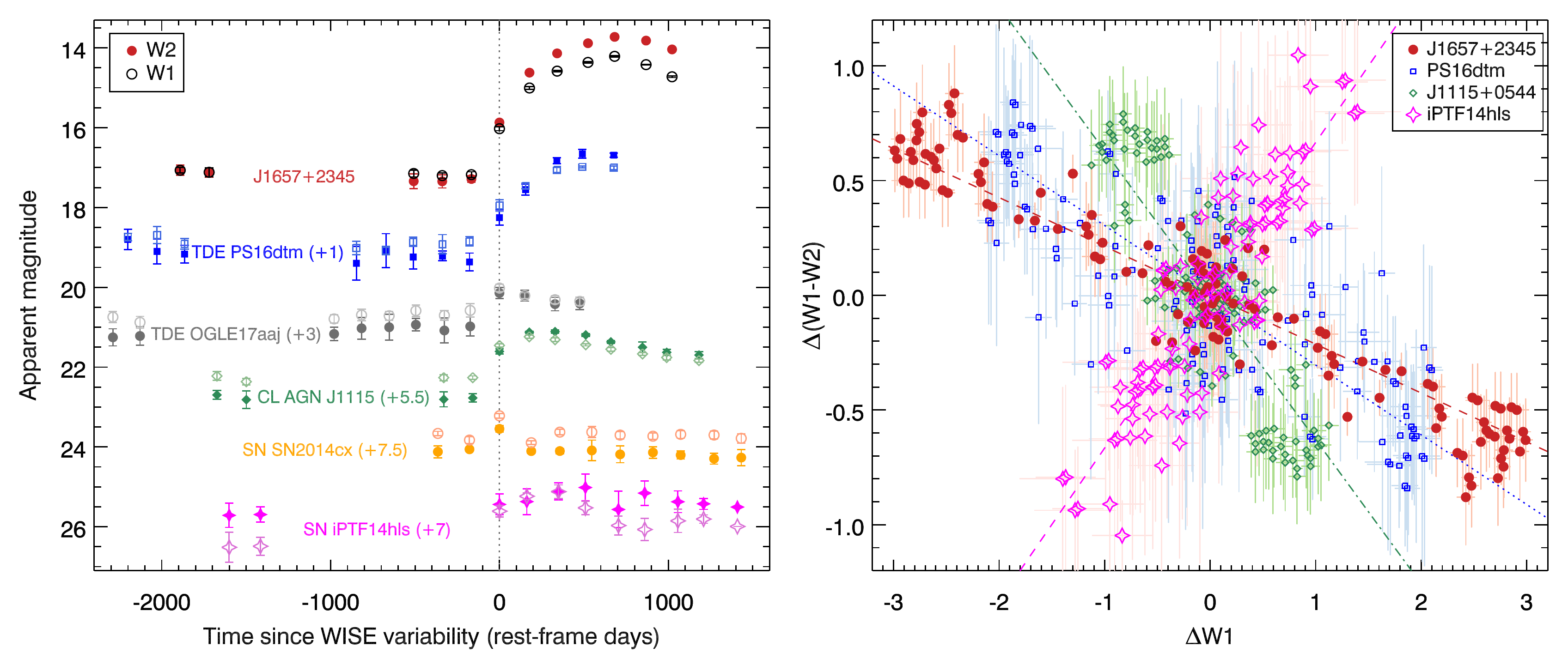}
\caption{\textbf{Left panel:} MIR light curves of J1657$+$2345 compared with several known transients. Filled and open shapes represent $W2$ and $W1$ bands, respectively. From top to bottom: J1657$+$2345 (red), TDE PS16dtm \citep[blue,][]{Blanchard2017}, TDE OGLE17aaj \citep[gray,][]{Gromadzki2019}, CL AGN J1115$+$0544 \citep[green,][]{Yang2018}, SN SN2014cx \citep[orange,][]{Guillochon2017}, and SN iPTF14hls \citep[magenta,][]{Arcavi2017}. \textbf{Right panel:} MIR color variability vs. magnitude variability. J1657$+$2345 (red filled circles), TDE PS16dtm (blue open squares), and CL AGN J1115$+$0544 (green open diamonds) are redder-when-brighter due to stronger hot dust emission in bright states, while SN iPTF14hls (magenta open stars) is bluer-when-brighter, possibly as a result of no hot dust emission. \label{fig:MIR} }
\end{figure*}

The results show that the host galaxy is dominated by the old stellar population, with stellar mass $M_{\rm *,old} = 10^{9.59\pm0.02}~M_\odot$ and age $1.45\pm0.04$ Gyr (summarized in Table \ref{tab:fit}). Using the relation between stellar mass and black hole mass of AGN in the nearby Universe \citep{Reines2015}, we estimate a black hole mass of $10^{5.97\pm0.30}~M_\odot$, {consistent with our earlier estimation using the $M-\sigma_*$ relation. }

To estimate the bolometric luminosity, $L_{\rm bol}$, of this Type 2 AGN, we first estimate the 6 \micron\ luminosity, $\lambda L_\lambda(6\,\micron)$, based on our SED fit. We then adopt $L(\mbox{2--10 keV})/\lambda L_\lambda(6\,\micron)=0.234$ \citep{Lutz2004} and the 2--10 keV bolometric correction $L_\mathrm{bol}/L(\mbox{2--10 keV})=20$ \citep[e.g.,][]{Elvis1994}. For the faint state, we obtained $\lambda L_\lambda(6\,\micron)=(3.0\pm1.9)\times10^{42}\,\mathrm{erg\,s^{-1}}$, and derived $L_\mathrm{bol,faint}=(1.4\pm0.9)\times10^{43}\,\mathrm{erg\,s^{-1}}$. J1657$+$2345 was not detected by the ROSAT All Sky Survey \citep{Boller2016}. Assuming a power-law in X-ray with a typical photon index of 2, we derive a flux (0.1-2.4keV) of $9.1\times10^{-14}\,\mathrm{erg\,s^{-1}\,cm^{-2}}$, which is below the ROSAT flux limit of a few times $10^{-13}\,\mathrm{erg\,s^{-1}\,cm^{-2}}$.
For the bright state, we got $\lambda L_\lambda(6\,\micron)=(1.0\pm0.1)\times10^{44}\,
\mathrm{erg\,s^{-1}}$, and derived $L_\mathrm{bol,bright}=(4.9\pm0.1)\times10^{44}\,\mathrm{erg\,s^{-1}}$. To estimate the peak bolometric luminosity, we use the {\it WISE} data in 2017 August, as well as the DBSP spectrum and UKIRT photometry, to derive $\lambda L_\lambda(6\,\micron)=(1.6\pm0.2)\times10^{44}\,
\mathrm{erg\,s^{-1}}$, thus $L_\mathrm{bol, peak}=(7.5\pm0.1)\times10^{44}\,\mathrm{erg\,s^{-1}}$.


\subsection{Spectral Properties \label{sec:spec_prop}}

To obtain emission line properties, we fit the optical spectra with stellar emission, described in Section \ref{sec:sed}, any residual (or scattered) power-law continuum from the AGN\footnote{However, this residual AGN continuum component is very weak and cannot be well constrained from the spectral fitting (see discussion at the end of \S\ref{sec:spec_prop}). }, broad-line emission, and narrow line emission. Figure \ref{fig:spec} shows an example of the spectral fitting to the SDSS and DBSP spectra. We summarize some spectral fitting properties in Table \ref{tab:fit}. We only show fitting to the SDSS and DBSP spectra, because the resolution and $S/N$ of the spectrum taken by Xinglong 2.16m telescope are lower and there is no measurable variation between this spectrum and the DBSP spectrum (see Table \ref{tab:Optical}).

The [\ion{O}{3}]$\lambda$5007 luminosity, $L_{\rm [OIII]}$, from the SDSS and DBSP spectra are $10^{40.45\pm0.01}$ and $10^{40.66\pm0.01}\,{\rm erg\,s^{-1}}$, respectively. We have scaled the $L_{\rm [OIII]}$ obtained from the spectra by a factor of 1.95 to correct for aperture losses (described in Section \ref{sec:spec}). We estimate the bolometric luminosity from $L_{\rm [OIII]}$ following \citet{Trump2015}:
\begin{equation}
    \frac{L_{\rm bol}}{10^{40}~{\rm erg~s}^{-1}} = 112 \left( \frac{L{\rm [O~III]}}{10^{40}~{\rm erg~s}^{-1}} \right)^{1.2}.
\end{equation}
Thus the SDSS and DBSP $L_{\rm [OIII]}$ corresponds to $L_{\rm bol, faint}=(7.57\pm0.01)\times10^{42}$ and $(1.35\pm0.01)\times10^{43}\,{\rm erg\,s^{-1}}$, which are both consistent with that obtained from the faint-state SED in Section \ref{sec:sed}. The [\ion{O}{3}]$\lambda$5007 luminosity in the bright state has not yet responded to the flare, as expected from the $\sim$kpc distances of the narrow-line region gas. 


There is weak broad H$\alpha$ emission detected in both the SDSS and the DBSP spectra. We have tested different fitting recipes for the narrow emission lines and different extraction apertures for the DBSP spectrum and found the broad H$\alpha$ measurements are robust against these details. The broad-line H$\alpha$ luminosity, $L_{\rm H\alpha,broad}$, is $10^{39.86\pm0.06}$ ($10^{40.53\pm0.05}$) $\mathrm{erg\,s^{-1}}$ from the SDSS (DBSP) spectrum, and the FWHM is $1610\pm210$ ($1664\pm161$) km\,s$^{-1}$. On the other hand, we do not detect broad H$\beta$ emission in the SDSS and DBSP spectra. 
The expected unobscured broad H$\alpha$ luminosity, using the measurements for broad-line AGN in \citet{Shen2011}, is about 1.5 dex larger than the \OIII\ luminosity (at $L_{\rm [OIII]}\sim 10^{40.5}\,{\rm erg\,s^{-1}}$) in the faint state. Thus we estimate an unobscured broad H$\alpha$ luminosity of $\sim 10^{42}\,{\rm erg\,s^{-1}}$. The detected broad H$\alpha$ flux in the faint state is then roughly 1\% of the unobscured broad line flux. Therefore the detected broad H$\alpha$ flux is consistent with scattered light in Type 2 AGN \citep[e.g.,][]{Zakamska2005}. If this is the case, it is reasonable to see an increase in the broad H$\alpha$ flux from the faint state to the bright state. The broad H$\alpha$ flux of the DBSP spectrum (bright state) is a factor of $\sim$5 of that from the SDSS spectrum (faint state). However, the MIR luminosity increased by a factor of $\sim$20 between the faint and bright states. The UV/optical flare could last much shorter than the MIR flare (see \S\ref{sec:torus_model}). Therefore the broad H$\alpha$ flux may have faded significantly when we took the DBSP spectrum.

The width and luminosity of the broad H$\alpha$ line can be used to estimate the AGN black hole mass \citep{Greene2005}. Using the detected scattered broad H$\alpha$ in the SDSS and DBSP spectra, and assuming a scattering fraction of 1\% to obtain the unobscured broad H$\alpha$ luminosity, we estimate a black hole mass of $10^{6.66\pm0.15}$ and $10^{7.05\pm0.10}~M_\odot$, which are slightly higher than the black hole masses estimated from stellar mass and $\sigma_*$, but broadly consistent given uncertainties in the scattered light fraction and in these BH mass proxies.



The $M_{\rm BH}$ estimates range from $10^{5.97}$ to $10^{7.05}$ using stellar mass, velocity dispersion, and scattered broad H$\alpha$ flux. We estimate the Eddington ratio $\lambda_{\rm Edd} = L_{\rm bol}/L_{\rm Edd}$, where $L_{\rm Edd}=1.38\times10^{38}(M_{\rm BH}/M_\odot)$. Using the average value of BH mass ($\sim 10^{6.5}\,M_\odot$) and bolometric luminosity to mitigate uncertainties in individual estimates, we estimate $\lambda_{\rm Edd}\sim 0.02$ for the faint state, $\lambda_{\rm Edd}\sim 1.1$ for the bright state, and $\lambda_{\rm Edd}\sim 1.7$ for the brightest (peak {\it WISE} flux) state.

The optical continuum is dominated by stellar emission given the Type 2 AGN nature. Using a typical bolometric correction of 10 for AGN continuum luminosity at 5100 ${\rm \AA}$, $\lambda L_{\lambda}(5100\,{\rm \AA})$, for unobscured quasars \citep[][]{Shen2011}, and assuming the same scattering fraction of 1\% as for broad H$\alpha$, a bolometric luminosity $\rm L_{bol,faint} \simeq 1.4\times10^{43}\,{\rm erg\,s^{-1}}$ corresponds to a scattered AGN $\lambda L_{\lambda}(5100\,{\rm \AA})$ of $1.4\times10^{40}\,{\rm erg\,s^{-1}}$. This is less than 1\% of the total observed continuum at 5100 ${\rm \AA}$. Thus, even if the obscured optical AGN continuum brightened by a factor of $\sim20$, the observed total flux can only be increased by less than 0.2 mag in $r$-band, consistent with observations.



\section{Discussion} \label{sec:disscussion}
\subsection{MIR Variability of Transients}
We compare the MIR variability of J1657$+$2345 with some known transient classes displaying MIR variability, including CL AGN, TDEs, and SNe. We show several examples of light curves in Figure \ref{fig:MIR}. 

\citet{Yang2018} found that many CL AGN exhibit MIR variability. They are redder-when-brighter due to stronger hot dust contribution in the $W2$ band when the AGN activity becomes stronger. 

Some studies \citep{Dou2016, Dou2017, Jiang2016, vanVelzen2016, Jiang2017} reported MIR transient events following a candidate TDE flare in the optical, which were interpreted as signatures of nuclear dust reprocessing the UV/optical flare. We checked the {\it WISE} MIR light curves of additional TDEs discovered since 2014, when {\it NEOWISE-R} began continuously scanning the full sky every six months. We found that one TDE OGLE17aaj, discovered by \citet{Gromadzki2019}, also displayed a MIR flare. No MIR flare was detected (at $>1\sigma$ significance) by {\it WISE} for the other TDEs, including ASASSN-14ae discovered by \citet{Holoien2014}, ASASSN-15oi discovered by \citet{Holoien2016}, OGLE16aaa discovered by \citet{Wyrzykowski2017}, iPTF16axa discovered by \citet{Hung2017}, and PS18kh discovered by \citet{Holoien2018}. 


We study the MIR variability of 2,812 SNe discovered from January, 2014 to September, 2018 from an open catalog for SNe \citep{Guillochon2017}. $\sim$ 58\% of them or their host galaxies were detected by {\it WISE}. {837 (30\%) of them were detected in more than one epochs by {\it WISE}.} Among them, 115 (36) had larger than 0.5 (1) mag variability in $W1$ band. SNe usually brighten in MIR for a shorter timescale (typically caught in one {\it WISE} epoch, i.e. less than one year) than CL AGN and TDEs (see an example of SN2014cx in the left panel of Figure \ref{fig:MIR}). iPTF14hls is a peculiar SN with long-term (a few years) MIR variability after the explosion, which was classified as type II-P SN and interpreted as a hydrogen-rich explosion of a massive star \citep{Arcavi2017}.

However, the evolution of MIR color of SNs are generally different from those of CL AGN and TDEs. Figure \ref{fig:MIR} (right) displays the dependence of color variability $\Delta(W1-W2)$ on the magnitude variability $\Delta W1$ for different populations. We fit a least-squares regression to $\Delta(W1-W2)$ and $\Delta W1$: \\
\begin{equation}
    \Delta (W1-W2) = {\rm A} \Delta W1.
\end{equation}
We obtain $\rm A=0.433\pm0.005$ for the {837 SNe detected in more than one epochs by {\it WISE}}, and $\rm A=0.666\pm0.045$ for iPTF14hls. On the other hand, we obtain $\rm A=-0.213\pm0.018$ for J1657$+$2345, $\rm A=-0.304\pm0.022~(-0.514\pm0.042)$ for TDE PS16dtm (OGLE16aaa), and $\rm A=-0.629\pm0.064$ for CL AGN J1115$+$0544. CL AGN and TDEs (PS16dtm and OGLE16aaa) all display a redder-when-brighter behavior, as a consequence of stronger hot dust radiation from nuclear dust torus that peaks at wavelengths redder than $W2$ band. SNe, however, displays a bluer-when-brighter behavior, possibly caused by the lack of hot dust contribution. 

J1657$+$2345 is redder-when-brighter in MIR, with emerging strong hot dust contribution in the bright state. Thus, we speculate that the J1657$+$2345 MIR transient is from the dust torus of the central AGN. It is likely that the central black hole of J1657$+$2345 is increasing its accretion rate as in a CL AGN, or due to a recent TDE embedded in a highly-obscuring dust torus. The flare of J1657$+$2345 is most likely due to TDE or CL AGN events. It is less likely that the MIR flare of J1657$+$2345 is due to a normal SN. The explosion of a massive star in the central AGN dust torus still remains a possibility based on the color variability argument. However, the detection of enhanced scattered broad H$\alpha$ emission and the observed large MIR luminosity of the flare are difficult to explain with stellar explosions. 


\subsection{Variability Timescales}\label{sec:torus_model}

The MIR flare of J1657$+$2345 lasts more than three years. The dust torus surrounding the central SMBH responds to continuum variations and re-radiate in the infrared. The continuum light arrives at different parts of the reverberating torus region at different times, and the reprocessed light reaches the observer at different later times. Therefore, the extended torus structure can lead to smoothed and stretched MIR light curves than the driving UV/optical light curve. 

\begin{figure}[htbp]
\hspace*{-0.5cm}
\epsscale{1.1}
\plotone{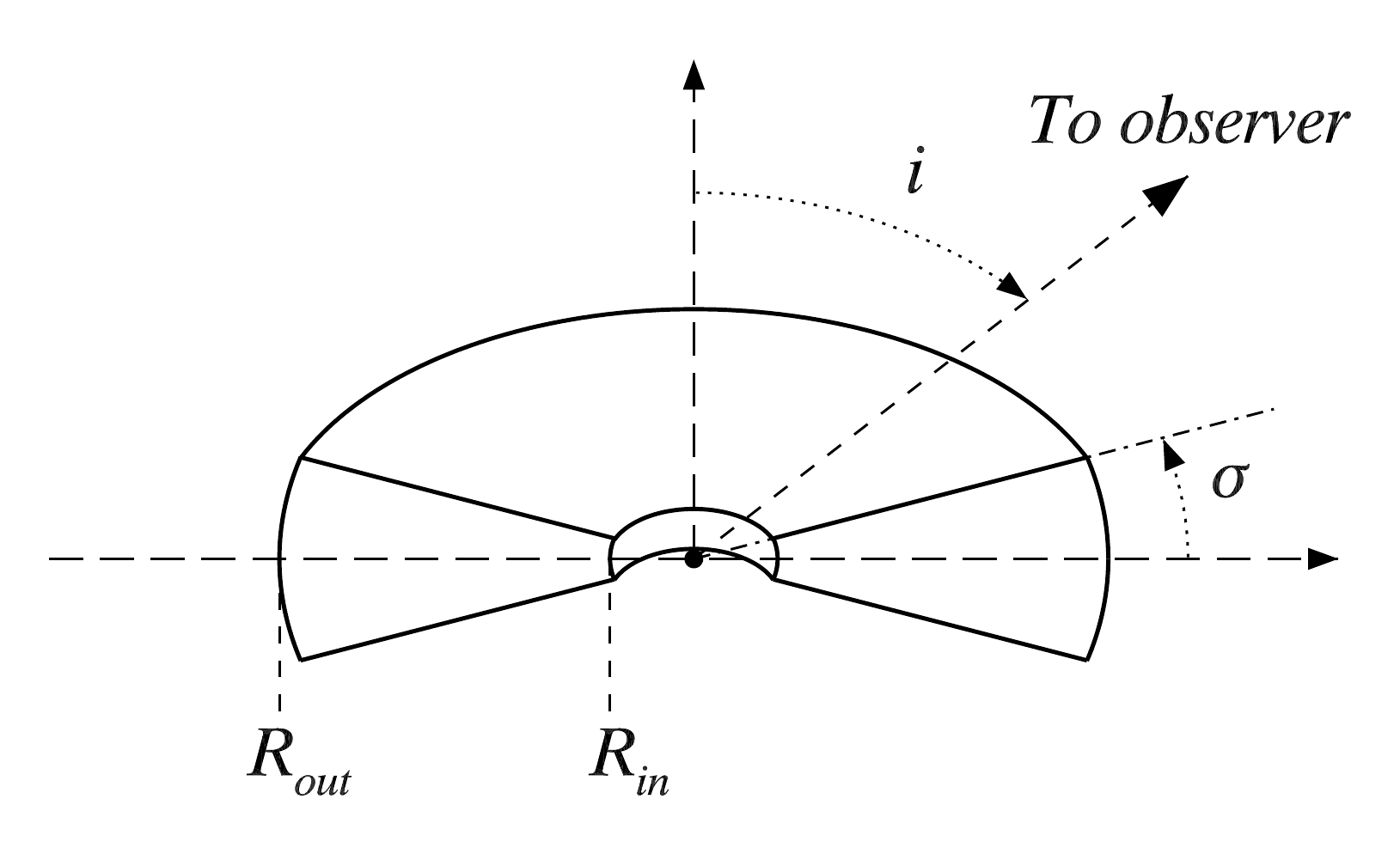}
\caption{A simple cartoon of the dust torus geometry \citep[e.g.,][]{Nenkova2008}. The inner and outer radii are $R_{in}$ and $R_{out}$. $\sigma$ is the half-opening angle and $i$ is the inclination angle. \label{fig:cartoon} }
\end{figure}

Here we construct a simple geometrical torus model (illustrated in Figure \ref{fig:cartoon}) to demonstrate the geometric effect of the dust torus on the IR echo. In our toy model, the dusty clouds are distributed between inner and outer radii $R_{in}$ and $R_{out}$, with half opening angle $\sigma$ and inclination angle $i$. Following the methodology in \citet{Shen2012}, we describe the driving UV/optical continuum flare as a step function at time $t = 0$ followed by a constant flux increment $f_c$ for a period of $\Delta t$ \citep[see eqn.~2 in][]{Shen2012}. We assume a constant density and uniform reprocessing efficiency across the entire torus region.

\begin{figure}[htbp]
\hspace*{-0.5cm}
\epsscale{1.1}
\plotone{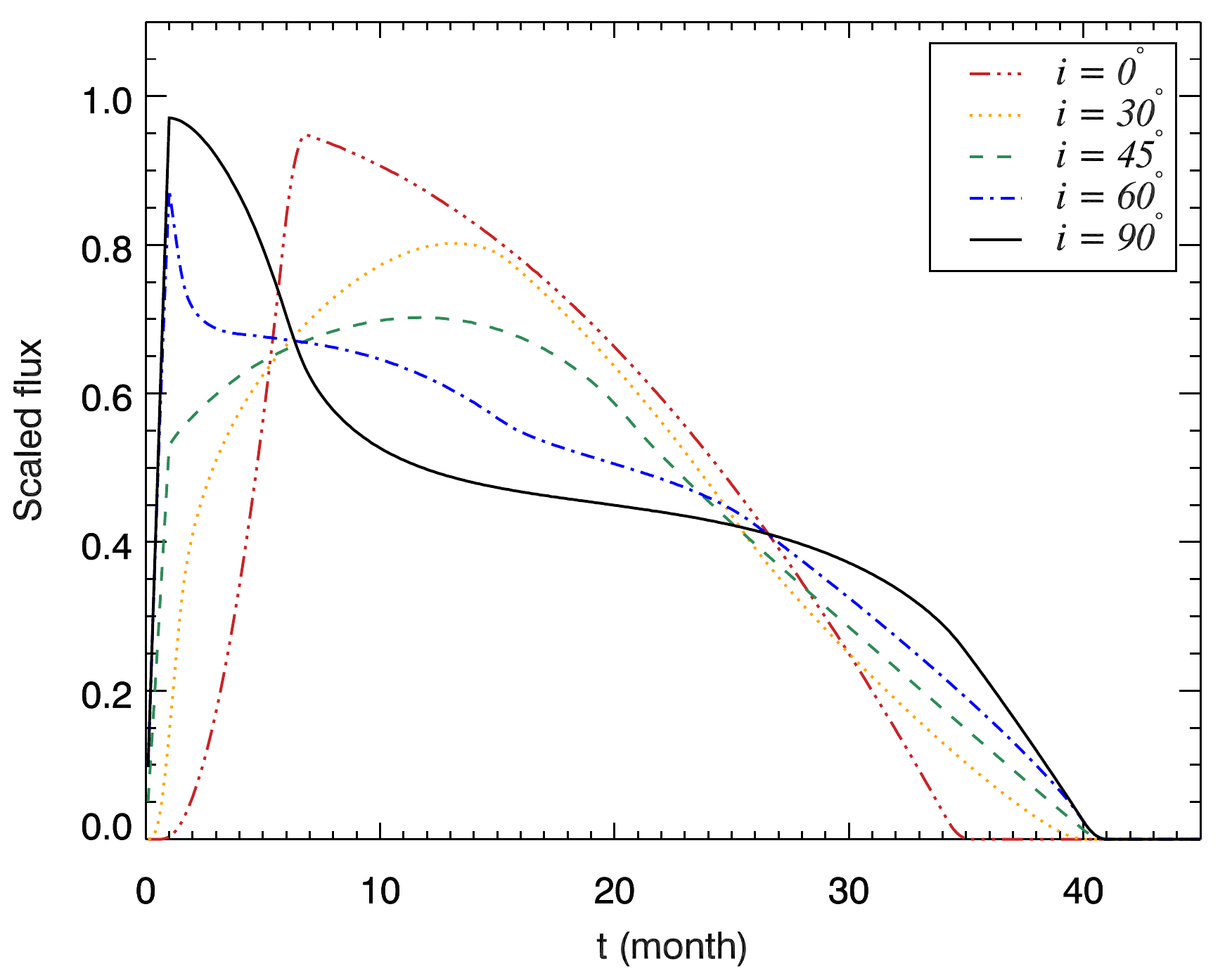}
\caption{Torus response curves at different inclination angle $i$ of 0$^\circ$ (red, i.e., face-on), 30$^\circ$ (orange), 45$^\circ$ (green), 60$^\circ$ (blue), and 90$^\circ$ (black, i.e., edge-on). The driving UV/optical flare is approximated by a constant flux of $f_c$ between $t=0$ and $t=1$ months and zero elsewhere. The responding light curves are much more extended as a consequence of the dust torus extension, and the inclination angle. \label{fig:LC_torus} }
\end{figure}

Figure \ref{fig:LC_torus} shows several examples of the responding MIR light curves for a dusty torus with $\sigma=45^\circ$, $R_{in}=2$, $R_{out}=20$, to a step function UV/optical flare with a duration $\Delta t=1$. Time is in units of month and distances are in units of light-month. For such a short UV/optical flare, the resulting MIR echo can be extended to $\sim$ three years due to the spatial extension of the dust torus. The responding MIR light curves also have different shapes for different inclination angles.

The inner radius of the torus is determined by the dust sublimation temperature as \citep{Nenkova2008} \\
\begin{equation}
    R_{in} \simeq 0.4 \left(\frac{L_{\rm bol}}{10^{45}~{\rm erg^{-1}}}\right)^{1/2}\left(\frac{1500~{\rm K}}{T_{\rm sub}}\right)^{2.6}~{\rm pc},
\end{equation}
where $T_{\rm sub}$ is the dust sublimation temperature and $L_{\rm bol}$ is the bolometric luminosity of the AGN. The innermost dust radius has now been measured by reverberation lags between optical and near-infrared ($K$-band) for $\sim$20 nearby AGN \citep[e.g.,][]{Minezaki2004, Suganuma2006, Koshida2014}. The inner dust radii of these AGN range from 10 to 150 light days, and correlate tightly with AGN luminosity, $L_{\rm AGN}$, as $R_{in} \propto L_{\rm AGN}^{0.5}$. It is more difficult to determine the outer boundary of the dust torus. Current observations are consistent with a torus radial thickness $Y = R_{out}/R_{in}$ no more than $\sim 20-30$, and perhaps $5-10$ \citep{Nenkova2008}. Thus the outer radius is a few light years.

Using the bolometric luminosity $L_{\rm bol,faint} = 1.35\times10^{43}~{\rm erg~s}^{-1}$ at the faint state, and assuming $T_{\rm sub} = 1500$ K, we obtain $R_{in} \sim 0.05$ pc, corresponding to 55 light days. A torus radial thickness $Y = 5-30$ translates to an outer boundary $R_{out}$ of approximately 0.8 to 4.6 light years. These estimates are similar to the values adopted for our demonstration example described earlier. 



J1657+2345 has an estimated BH mass of $\sim 10^6-10^7\,M_\odot$ (Table \ref{tab:fit}). For a $M_{\rm B H} = 10^{6}~M_\odot$ SMBH, the Schwarzschild radius is $R_S = 2GM_{\rm BH}/c^2 = 3\times10^{11}$ cm. The tidal disruption radius is $r_{\rm T} \simeq 5\times10^{12} M_6^{1/3}(r_*/r_\odot)(m_*/M_\odot)^{-1/3}$ cm, where $M_6 = M_{\rm BH}/10^6~M_\odot$, $m_*$ and $r_*$ are the mass and radius of the disrupted star \citep{Rees1988}. Thus the tidal disruption radius of solar-type stars is much larger than the Schwarzschild radius for a $M_{\rm} = 10^{6}~M_\odot$ SMBH. The characteristic timescale of TDE, i.e., the orbital period of the most tightly bound debris, is $\Delta t = 0.35 M_7^{1/2}(m_*/M_\odot)^{-1}(r_*/r_\odot)^{3/2}\simeq 0.1 $ yr, where $M_7 = M_{\rm BH}/10^7~M_\odot$ \citep{Lodato2011}. Therefore, the observed multi-year MIR light curve for J$1657+2345$ could be the stretched response to the much shorter TDE UV/optical flare of a few months. Indeed, \citet{Mattila2018} reported a dust-enshrouded TDE in a nearby merging galaxy Arp 299, whose MIR light curves are similar to that of J1657$+$2345. The low BH mass and the approximately Eddington-limited luminosity near the peak of the light curve makes a TDE a favorable scenario for the flare in J1657$+$2345.


Rare, rapid CL AGN phenomena have also been observed on timescales of less than 1 year \citep{Gezari2017, Yang2018, Yan2019, Trakhtenbrot2019b}. Thus the MIR flare of J$1657+2345$ could also be the stretched dust echo of such rapid CL AGN events. While in general CL AGN are not TDEs, some of them may be due to TDEs with observed similar decaying light curves. Indeed, the TDE scenario has been invoked to explain specific CL AGN \citep[e.g.,][]{Merloni2015}. However, the dramatic changes in the accretion flow onto the SMBH are still not fully understood \citep[e.g.,][]{Rumbaugh2018,Trakhtenbrot2019,Dexter_etal_2019}, and it is possible that other processes can drive the required UV/optical variability without tidally disrupting a star. 



\section{Summary} \label{sec:summary}

We have discovered an exceptional MIR transient in the Type 2 AGN J1657$+$2345, with its {\it WISE} MIR magnitudes brightened by 3 mag from 2015 to 2017. Among more than 1.8 million galaxies in the SDSS DR14, J1657$+$2345 has the most extreme {\it WISE} variability amplitude. MIR variability is commonly interpreted as the dust echo of UV/optical variability. However, for this peculiar MIR transient, there is no corresponding optical photometric variability from 2003 to 2019. New optical spectra confirm that there is no significant optical variability, and new near-infrared photometry demonstrates that the MIR flare is more prominent at longer wavelengths. The lack of optical variability distinguishes J1657$+$2345 from the known population of CL AGN. 

We modeled the SEDs and optical spectra of J1657$+$2345, and estimated a black mass of $\sim 10^{6}-10^{7}~M_\odot$ and stellar mass $\sim 10^{9.6}~M_\odot$. The bolometric luminosity in the faint state is $1.4\times10^{43}$ erg s$^{-1}$ and in the most recent bright state is $4.9\times10^{44}$ erg s$^{-1}$. The peak bolometric luminosity is $7.7\times10^{44}$ erg s$^{-1}$. These estimates place the accretion luminosity of the BH around the Eddington limit in the bright state. Based on energetic grounds and the multi-year long duration of the flare, it would be difficult to associate the MIR flare with stellar explosions and circumstellar dust reprocessing. In addition, the MIR color variability is consistent with CL AGN and TDEs, and inconsistent with normal SNe. We also detected weak broad H$\alpha$ emission in the faint and bright-state spectra, which is consistent with scattered light from the obscured broad-line region. The enhancement of the scattered broad H$\alpha$ emission in the latest spectrum also favors the AGN variability and torus reprocessing scenario. 

Assuming that the MIR flare comes from the AGN torus reprocessing the UV/optical variability from much closer to the BH, we show that the extended torus geometry can reproduce the multi-year MIR light curve responding to a nuclear UV/optical flare on much shorter timescale of a few months. 

Overall our best explanation for the J1657$+$2345 MIR flare is  that it is a rapidly turning-on AGN or TDE that is heavily obscured in the optical in a Type 2 AGN. Such objects are naturally expected as the Type 2 counterparts to the optically unobscured CL AGN population. The discovery of the J1657$+$2345 flare reveals there is a population of similar nuclear flares that are missed in the optical window. A systematic search of such MIR transients among different types of galaxies will be important to understand nuclear transients in general. 




\acknowledgments

QY and YS acknowledge support from NSF grant AST-1715579. XW, LJ, and SY acknowledge support from the National Key R\&D Program of China (2016YFA0400703) and the National Science Foundation of China (11533001, 11721303, 11890693). MG acknowledges support from NSF grant AST-1815034.

This publication makes use of data products from the {\it WISE}, which is a joint project of the University of California, Los Angeles, and the Jet Propulsion Laboratory/California Institute of Technology, funded by the National Aeronautics and Space Administration. This publication also makes use of data products from {\it NEOWISE}, which is a project of the Jet Propulsion Laboratory/California Institute of Technology, funded by the Planetary Science Division of the National Aeronautics and Space Administration.

Funding for the SDSS and SDSS-II has been provided by the Alfred P. Sloan Foundation, the Participating Institutions, the National Science Foundation, the U.S. Department of Energy, the National Aeronautics and Space Administration, the Japanese Monbukagakusho, the Max Planck Society, and the Higher Education Funding Council for England. The SDSS Web Site is http://www.sdss.org/. The SDSS is managed by the Astrophysical Research Consortium for the Participating Institutions. The Participating Institutions are the American Museum of Natural History, Astrophysical Institute Potsdam, University of Basel, University of Cambridge, Case Western Reserve University, University of Chicago, Drexel University, Fermilab, the Institute for Advanced Study, the Japan Participation Group, Johns Hopkins University, the Joint Institute for Nuclear Astrophysics, the Kavli Institute for Particle Astrophysics and Cosmology, the Korean Scientist Group, the Chinese Academy of Sciences (LAMOST), Los Alamos National Laboratory, the Max-Planck-Institute for Astronomy (MPIA), the Max-Planck-Institute for Astrophysics (MPA), New Mexico State University, Ohio State University, University of Pittsburgh, University of Portsmouth, Princeton University, the United States Naval Observatory, and the University of Washington.

The PS1 and the PS1 public science archive have been made possible through contributions by the Institute for Astronomy, the University of Hawaii, the Pan-STARRS Project Office, the Max-Planck Society and its participating institutes, the Max Planck Institute for Astronomy, Heidelberg and the Max Planck Institute for Extraterrestrial Physics, Garching, The Johns Hopkins University, Durham University, the University of Edinburgh, the Queen's University Belfast, the Harvard-Smithsonian Center for Astrophysics, the Las Cumbres Observatory Global Telescope Network Incorporated, the National Central University of Taiwan, the Space Telescope Science Institute, the National Aeronautics and Space Administration under Grant No. NNX08AR22G issued through the Planetary Science Division of the NASA Science Mission Directorate, the National Science Foundation Grant No. AST-1238877, the University of Maryland, Eotvos Lorand University (ELTE), the Los Alamos National Laboratory, and the Gordon and Betty Moore Foundation.

The Legacy Surveys consist of three individual and complementary projects: the Dark Energy Camera Legacy Survey (DECaLS; NOAO Proposal ID 2014B-0404; PIs: David Schlegel and Arjun Dey), the Beijing-Arizona Sky Survey (BASS; NOAO Proposal ID 2015A-0801; PIs: Zhou Xu and Xiaohui Fan), and the Mayall z-band Legacy Survey (MzLS; NOAO Proposal ID 2016A-0453; PI: Arjun Dey). DECaLS, BASS and MzLS together include data obtained, respectively, at the Blanco telescope, Cerro Tololo Inter-American Observatory, National Optical Astronomy Observatory (NOAO); the Bok telescope, Steward Observatory, University of Arizona; and the Mayall telescope, Kitt Peak National Observatory, NOAO. The Legacy Surveys project is honored to be permitted to conduct astronomical research on Iolkam Du'ag (Kitt Peak), a mountain with particular significance to the Tohono O'odham Nation. NOAO is operated by the Association of Universities for Research in Astronomy (AURA) under a cooperative agreement with the National Science Foundation.

The CSS survey is funded by the National Aeronautics and Space
Administration under Grant No. NNG05GF22G issued through the Science
Mission Directorate Near-Earth Objects Observations Program.  The CRTS survey is supported by the U.S.~National Science Foundation under grants AST-0909182.

ZTF: Based on observations obtained with the Samuel Oschin 48-inch Telescope at the Palomar Observatory as part of the Zwicky Transient Facility project. ZTF is supported by the National Science Foundation under Grant No. AST-1440341 and a collaboration including Caltech, IPAC, the Weizmann Institute for Science, the Oskar Klein Center at Stockholm University, the University of Maryland, the University of Washington, Deutsches Elektronen-Synchrotron and Humboldt University, Los Alamos National Laboratories, the TANGO Consortium of Taiwan, the University of Wisconsin at Milwaukee, and Lawrence Berkeley National Laboratories. Operations are conducted by COO, IPAC, and UW. 

This publication makes use of data products from the 2MASS, which is a joint project of the University of Massachusetts and the Infrared Processing and Analysis Center/California Institute of Technology, funded by the National Aeronautics and Space Administration and the National Science Foundation.


\bibliography{references.bib}

\end{document}